\documentclass[apjl]{emulateapj}

\usepackage{amsmath}
\usepackage{apjfonts}
\usepackage{amssymb}
\usepackage{mathrsfs}
\usepackage{natbib}
\usepackage{color}
\def\citea#1{\citep{#1}}
\def\citea#1{}
\newcommand{\Msun}{\mbox{$M_\odot$}}

\begin{document}

\shorttitle{Mergers of BHs in Astrophysical Environments}

\shortauthors{Bode et al.}

\title{Mergers of Supermassive Black Holes in Astrophysical Environments}

\author{Tanja Bode\altaffilmark{1}, Tamara
Bogdanovi\'c\altaffilmark{2,3}, Roland Haas\altaffilmark{1}, James
Healy\altaffilmark{1},\\ Pablo Laguna\altaffilmark{1}\& Deirdre
Shoemaker\altaffilmark{1}}

\altaffiltext{1}{Center for Relativistic Astrophysics and School of
  Physics, School of Physics, Georgia Institute of Technology,
  Atlanta, GA 30332, USA} 
\altaffiltext{2}{Department of Astronomy, University of Maryland, College Park, MD 20742, USA} 
\altaffiltext{3}{Einstein Postdoctoral Fellow}

\begin{abstract}

Modeling the late inspiral and merger of supermassive black holes is
central to understanding accretion processes and the conditions under
which electromagnetic emission accompanies gravitational waves. We use
fully general relativistic, hydrodynamics simulations to investigate
how electromagnetic signatures correlate with black hole spins, mass
ratios, and the gaseous environment in this final phase of binary
evolution. In all scenarios, we find some form of characteristic
electromagnetic variability whose detailed pattern depends on the
spins and binary mass ratios. Binaries in hot accretion flows exhibit
a flare followed by a sudden drop in luminosity associated with the
plunge and merger, as well as quasi-periodic oscillations correlated
with the gravitational waves during the inspiral. Conversely,
circumbinary disk systems are characterized by a low luminosity of
variable emission, suggesting challenging prospects for their
detection.
\end{abstract}


\maketitle

%%%%%%%%%%%%%%%%%%%%%%%%%%%%%%%%%%%%%%%%

\section{Introduction}\label{S_intro}

Gravitational waves (GWs) from the inspiral and coalescence of
supermassive binary black holes (BBHs) contain detailed information
about the black hole (BH) masses, their spins, and the orbital
characteristics of the binary. Space-based GW observations will
provide measurements of these quantities with a level of accuracy
rarely attained by astronomical observations
\citep{PhysRevD.77.024030,2009PhRvD..80f4027K}.  Electromagnetic (EM)
observations of these cataclysmic events, on the other hand, can
provide an observational link between the merging BHs and their host
galaxies by shedding light on the environment surrounding the holes
and, in particular, by giving us insight about accretion and feedback
processes. Our work investigates the circumstances under which
detectable EM emission accompanies the GW signal from supermassive BBH
mergers in astrophysical environments.

Our work complements non-relativistic hydrodynamic simulations that
follow the inspiral of a supermassive BH pair in the aftermath of
galactic merger down to a scale of $\sim 0.01$pc.  \citep[see][ for a
review]{colpi07}.  These studies have provided clues about BH
interactions with the surrounding stars and gas that lead to the
formation of gravitationally bound binaries. They suggest that
characteristic EM signatures may be associated with supermassive BBHs
in this stage of evolution \citep[][for example]{an02, mp05, dotti06,
bogdanovic08}.  Modeling at even smaller scales, where the orbital
dynamics of the binary is determined by gravitational radiation,
requires a post-Newtonian treatment of gravity and, in the last few
tens of orbits and merger, the framework of general relativity.
Examples of such works include recent numerical relativity studies of
coalescing binaries surrounded by test particles \citep{vanmeter09},
gas \citep{bode10, farris10}, or electromagnetic fields
\citep{palenzuela09,palenzuela10, mosta10}.

Previously, in \citet{bode10}, we carried out the first fully general
relativistic, hydrodynamics study of the late inspiral and merger of
binaries with equal-mass and parallel-spin BHs in a hot accretion
flow. This work together with \citet{farris10} established that plunge
and merger of ``symmetric'' binaries in hot accretion flows are
characterized by an EM flare followed by a sudden drop-off in
luminosity. Another of its findings is that the BBH orbital motion
imprints a quasi-periodic signal in the light curve that is an EM
equivalent of a GW chirp. However, it remained to be answered whether
such signals are present for more generic binary configurations and in
cooler accretion flows such as accretion disks. The present work
investigates the effect of varying BH spins and mass ratios and
considers astrophysically relevant environments that include
circumbinary disks and hot, radiatively inefficient accretion flows.
The hot flow and circumbinary disk effectively bracket a range of
physical scenarios for a BBH environment characterized by the balance
of heating and cooling processes in the gas. If cooling is more
efficient, the gas settles into a rotationally supported accretion
disk whereas when heating dominates it gives rise to a hot, tenuous,
and geometrically thick accretion flow.

We consider binaries with BH masses $m_i$ ($q = m_2/m_1 < 1$) and
dimensionless spin parameters $\vec{a}_i/m_i$. The initial binary is
on a quasi-circular orbit at a separation of $8\,M$ (in geometrical
units), with $M = m_1+m_2$ the total mass of the binary. We compute
our results using $M$ as a natural unit and present results scaled to
a total mass of $10^7\,M_\odot$; that is, $ M = 1.48\times
10^{12}\,{\rm cm}\;M_7 = 49.4\,{\rm s}\; M_7$ with $M_7 =
M/10^7\,M_{\odot}$. Since in all scenarios the mass of the gas in the
vicinity of the binary is many orders of magnitude lower than $M$, the
BBH dynamics is indistinguishable from the vacuum case.  We use the
computational methodology and infrastructure described in
\citet{bode10, Ansorg:2004ds, Baiotti:2004wn, Cactuscode.org:web,
EinsteinToolkit.org:web, Husa:2004ip, Schnetter:2003rb,
Thornburg:2003sf}.  Our simulations do not capture radiative
transport, magnetic fields nor physical viscosity.  Our computational
grid had an outer boundary at $\sim320\,M$, with 9 (10) levels of mesh
refinement for the equal (unequal) BH mass binaries. Resolution at the
finest level was $M/67 (M/76)$, with subsequent levels increased by a
factor of two.  The 5 (7) finest levels had $42^3 (36^3)$ grid points,
while remaining had $84^3 (72^3)$ grid points.  A subset of runs at
higher resolution of $M/100$ showed that our results did not change
significantly with resolution to affect our conclusions.

%%%%%%%%%%%%%%%%%%%%%%%%%%%%%%%%%%%%%%%%%%%%%%%%%%%%%%%%%%%%%%%%%%%%%%%%

\section{Hot Accretion Flow}\label{S_riaf}

In this scenario, the BBH environment is assumed to have the
properties of a radiatively inefficient accretion flow (RIAF). In
RIAFs, most of the energy generated by accretion and turbulent
stresses is stored as thermal energy in the gas, and the accretion
flow is hot and geometrically thick \citep{ichimaru77, ny94}. The
electron and ion plasmas in RIAFs can form a two-temperature flow in
which the thermal energy is stored in the ion plasma while the
electron plasma cools more efficiently (i.e., $T_p \gtrsim T_e$). In
such a case, the temperature of the plasma is represented by the ion
temperature, while the characteristics of emitted radiation depend on
the properties of electrons. The temperature ceiling reached by the
ion plasma will be determined by cooling processes such as the thermal
bremsstrahlung, synchrotron, and inverse Compton emission, as well as
the electron--positron pair production and the pion decay resulting
from energetic proton--proton collisions. Which process dominates the
energy loss of the plasma sensitively depends on its density,
temperature, and magnetic field strength, as well as the efficiency of
coupling between ions and electrons. The latter process determines the
rate with which energy can be transferred from hot ions to electrons,
and consequently the ratio of their temperatures,
$\varepsilon=T_e/T_p$.  While modeling of the radiative cooling and
treatment of the two-temperature plasma flow is beyond the
capabilities of our code at this time, we capture the effect of
different thermal properties of the plasma by investigating several
initial ion plasma temperatures in our simulations, $T_p = \lbrace
10^{10}, 10^{11}, 10^{12}\rbrace$K. In order to evaluate the emission
properties of these flows, we make a simplistic assumption that
$\varepsilon = 10^{-2}$ everywhere in the accretion flow. This is an
idealization as $T_e/T_p$ is expected to vary in both space and time
and can have a range of values between $\sim10^{-2}$ and 0.1 depending on
the dominant plasma processes \citep[see, for example,][]{sharma07}. Our
choice of $\varepsilon$ is however conservative because it caps $T_e$
and consequently the bremsstrahlung luminosity at lower values (see
\S~\ref{S_properties1}). The properties of the binary and initial
$T_p$ of the environment are listed in top part of
Table~\ref{tab:big}.

\subsection{Initial Conditions}

The gas around the binary is initialized with uniform density and
temperature and modeled with a polytropic index $\gamma=5/3$, adequate
for the ion plasma which is non-relativistic at temperatures
$T_p\lesssim10^{12}$K ($kT_p \lesssim 100\,$MeV). The gas initially
has zero linear and angular momentum. The latter condition is
consistent with the expectations based on the self-similar solutions
for RIAFs described by \citet{ny94} for $\gamma=5/3$ gas. This initial
configuration is first evolved on the static BBH spacetime for the
duration of $32\,M$ and subsequently on a dynamic BBH spacetime for
$160\,M$.\footnote{To ensure that this initial phase of relaxation
does not introduce spurious transients to our analysis, we omit it and
only report the subsequent evolution.} In comparison, the relaxation
time of the gas within the Bondi radius of the gravitational influence
of the binary ($t_\mathrm{relax} = R_\mathrm{B}/c_s$, where $c_s$ is
the speed of sound) is $18\,M$, $530\,M$, and $1.7\times 10^4\,M$ for
$10^{12}$ K, $10^{11}$ K and $10^{10}$ K gas, respectively.
\begin{table}[ht]
\centering
\begin{tabular}{l|ccc|cc} 
\hline\hline
  Case & $q$ & $\vec{a}_1/m_1$ & $\vec{a}_2/m_2$  & $T_{\rm p} \mathrm{(K)}$ \\
\tableline
  CH1& 1   & $(0,0,0.6)$    &  $(0,0,0.6)$   & $10^{12}$ \\ % Cloud A, G2
 CM1& 1   & $(0,0,0.6)$    & $(0,0,0.6)$    & $10^{11}$ \\ % Cloud B, G2
  CL1& 1   & $(0,0,0.6)$    & $(0,0,0.6)$    & $10^{10}$ \\ % Cloud C, G2
  CH2& 1/2 & $(0,0,0.6)$    & $(0,0,0.6)$    & $10^{12}$ \\ % Cloud A, G2q2
  CH3& 1/2 & $(-0.40,0.44,-0.02)$  & $(-0.16,0.54,-0.21)$ & $10^{12}$ \\ % Cloud A, short Generic. 
  CH4& 1/2 & $(-0.35,-0.47,0.10)$  & $(0.28,0.44,0.30)$ & $10^{12}$ \\ % Cloud A, long Generic.
  \tableline
 &  &  &  & $h/r$  \\
\tableline
  DA1& 1   & $(0,0,0.6)$     & $(0,0,0.6)$    & 0.2  \\ % Disk A, G2
  DB1& 1   & $(0,0,0.6)$    & $(0,0,0.6)$   & 0.4          \\ % Disk B, G2
  DC1$\,^*$& 1   & $(0,0,0.6)$     & $(0,0,0.6)$   &     0.2  \\   % Disk D, G2
  DA2& 1/2 & $(0,0,0.6)$     & $(0,0,0.6)$   & 0.2          \\ % Disk A, G2q2
  DA3& 1/2 & $(-0.40,0.44,-0.02)$  & $(-0.16,0.54,-0.21)$ & 0.2       \\ % Disk A, short Generic. 
  DA4& 1/2 & $(-0.35,-0.47,0.10)$  & $(0.28,0.44,0.30)$ & 0.2                  \\ % Disk A, long Generic. 
\hline\hline
\end{tabular}
\caption{ Parameters for the scenarios discussed for hot accretion flows (top half)
and circumbinary disks (bottom half).  The central columns describe the
the black hole parameters of the system while the last column designates
the primary parameter for the initial gas configuration.  The asterisk
next to DC1 refers to retrograde circumbinary disk rotation. }
\label{tab:big}
\end{table}
Longer relaxation time scales for $10^{10}$K and $10^{11}$K gas imply
that at the beginning of the simulation the gas is still settling into
a quasi-hydrostatic equilibrium and flowing toward the center of the
potential. We verified for all simulations that relaxation is a
gradual process which can be reproduced in simulations of accretion
flows with the same properties surrounding a single black hole with
mass equal to that of the BBH. We use the fact that the relaxation
process does not introduce rapid transients (i.e., on the BBH orbital
time scale) in the light curve of either system to remove a smooth
secular modulation in luminosity amplitude caused by relaxation by
dividing a light curve calculated for the BBH system by its single BH
equivalent. Using this ansatz we disentangle the effects of the gas
relaxation from variability driven by the orbiting BBH while at the
same time reducing the computational expense of our simulations to
about $800\,M$ per simulation. This computational gain ultimately
allows us to explore a relatively wide parameter space of the BBHs and
gas in this work.
\begin{figure*}[t]
\centering
\includegraphics[width=0.96\linewidth]{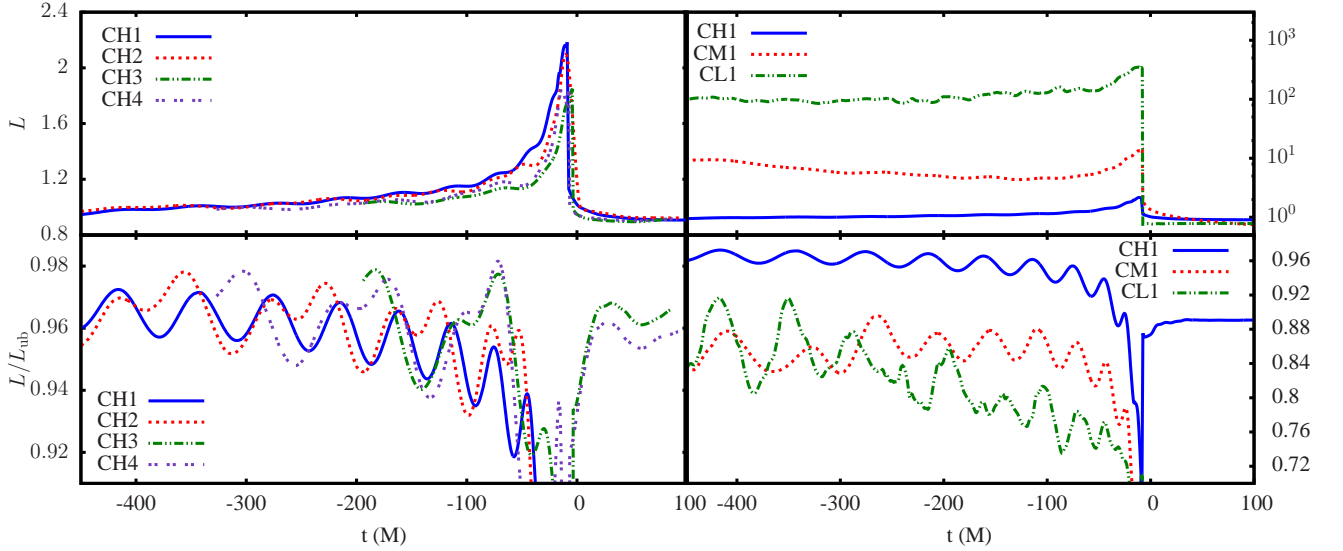}
% the plot* commands are used by the aastex target. Please do not remove them.
%\plotone{figures/Cloud_LBS-bw}\\
%\plotone{figures/Cloud_LBS-color}
\caption{{\it Top}: bremsstrahlung luminosity in hot accretion flows for
  different BBH configurations ({\it left}) and accretion flow
  temperatures ({\it right}) normalized to the luminosity of a single BH
  with corresponding mass.  {\it Bottom}: ratio of beamed to unbeamed
  luminosity $L_{\rm ub}$ for different BBH configurations ({\it left})
  and different accretion flow temperatures ({\it right}) for an observer
  viewing the binary edge-on.  Coalescence occurs at $t=0$. }
\label{fig:cloud_brems}
\end{figure*}

\subsection{Properties of the Accretion Flow}\label{S_properties1}

In all models, a pair of denser gas wakes quickly forms behind the
orbiting holes after the beginning of the simulation. About two orbits
before merger, a high-density, dynamically unstable region is also
formed inside the binary orbit.  After coalescence, this structure is
promptly swallowed by the newly formed BH. Both the wakes and the 
high-density central region are associated with strong shocks, driven by
the orbiting BBH, and their evolution gives rise to the characteristic
variability of the emitted light. This is illustrated in the top
panels of Figure~\ref{fig:cloud_brems}, which show bremsstrahlung
luminosity as a function of time, integrated within the sphere of
radius $R = 20M$ and normalized by the light curve for a single BH
with equivalent mass. We verified that variability in the accretion
flow associated with the BBH evolution is enclosed within this volume
by varying our choice of $R$ and confirmed that the central region
contributes the dominant component of the luminosity to the light
curve of the binary.

The top-left panel in Figure~\ref{fig:cloud_brems} shows a broad peak
in luminosity whose growth coincides with the formation of the shocked
region within the binary orbit. As the binary shrinks, the brightness
of this region increases until merger at which point the final BH
swallows the shocked gas and the luminosity precipitously drops off.
The top-left panel of Figure~\ref{fig:cloud_brems} compares the light
curves calculated for different BBH configurations in an environment
with initial temperature $T_p \approx 10^{12}$ K. The $q=1/2$ case
(CH2) exhibits a lower and narrower luminosity peak relative to the
$q=1$ case (CH1). This is because orbital torques from an unequal-mass
binary are less efficient at driving shocks in the gas, resulting in a
less luminous and delayed emission from this region.  In $q=1/2$
systems with generic spin orientations (CH3 and CH4), the luminosity
peaks at even lower values, at $\sim 80$\% of the amplitude achieved
by the $q=1$ binary. This is a consequence of the orbital precession
present in binaries with misaligned spins, which further inhibits the
formation of a stable shock region between the holes. The gradual rise
and sudden drop-off in luminosity however seem to be {\it a generic
feature of all modeled light curves, regardless of the spin
configuration and the mass ratios} explored in our simulations.
Moreover, the same feature has been observed for a wide range of
initial conditions employed in previous works carried out by our group
\citep{bode10} and by other authors \citep{farris10}, indicating that
this is a robust signature of binary systems merging in hot accretion
flows.

To understand the dependence of the luminosity on the properties of
the system, we estimate the bremsstrahlung luminosity emitted from the
Bondi radius of gravitational influence, $R_{\rm B}\approx
6.5\,M\,T_{p,12}^{-1}$, as
\begin{eqnarray}
 L_{\rm brem}  &\approx &  6.7\times 10^{40}\,{\rm erg\,s^{-1}} 
\varepsilon_{-2}^{1/2}\; T_{p,12}^{-1/2}\; \nonumber\\
& \times & \left(1 + 4.4\,\varepsilon_{-2}\;T_{p,12}\right)_{5.4}
\tau^2_{T}\; M_7^4 \,,
%  \varepsilon_{-2}^{1/2}\left(\frac{T_p}{10^{12}\,{\rm K}}\right)^{-1/2}\nonumber\\
%   & \times & \left[ 1 + 4.4\,\varepsilon_{-2}\left(\frac{T_p}{10^{12}\,{\rm K}}\right) \right]_{5.4}
%  \tau^2_{T}\; M_7^4 \,,
\label{eq_Lbrem}
\end{eqnarray}
where $\varepsilon = 10^{-2}\varepsilon_{-2}$, $T_p = 10^{12}{\rm K}\;
T_{p,12}$, $\tau_{T} = \sigma_{\rm T}\,\rho\,R_{\rm B}$ is the optical
depth to Thomson scattering within the Bondi sphere, $\sigma_{\rm T}$
is the cross section to Thomson scattering, and $\rho$ is the gas
density. Subscript ``5.4'' indicates that the expression in the
brackets is normalized to this value. Note that Eq. (\ref{eq_Lbrem})
implies {\it a maximum} bremsstrahlung luminosity that can be reached
by an accretion flow as long as its optical depth $\tau_{\rm
T}\lesssim 1$ (we consider Thomson scattering to be the dominant
source of opacity in this regime). Flow with a larger optical depth
would be subject to radiation pressure which could alter the
kinematics of the gas or unbind it from the BBH altogether, thus
acting to erase the variability imprinted by the binary motion and
suppress the luminosity.

The top-right panel in Figure~\ref{fig:cloud_brems} shows the light
curves calculated for parallel-spin $q=1$ binaries in gas with initial
temperatures $T_p = \lbrace 10^{10}, 10^{11}, 10^{12}\rbrace$ K. Lower
temperature flows tend to exhibit a larger luminosity peak and a more
dramatic drop-off than the hotter flows. This is because in colder
flows the shocks can be very effectively excited by the merging
binary. We find that regardless of its initial temperature, the gas in
the vicinity of the holes is persistently shock-heated to a
temperature $\sim10^{12}$ K as a consequence of the binary orbital
motion. Hence, the height of the peaks in the top-right panel of
Figure~\ref{fig:cloud_brems} reflects the ratio of bremsstrahlung
emissivities of the gas in the BBH system ($T_p \approx 10^{12}$ K) to
that in the single BH system ($T_p \approx \lbrace 10^{10}, 10^{11},
10^{12}\rbrace$ K), where bremsstrahlung emissivity $\propto
T_p^{1/2}(1 + 4.4\,\varepsilon_{-2}\; T_{p,12})$. It follows from this
simple estimate that the relative height of the BBH luminosity peak is
54, 12, and 1, respectively, consistent within a factor of few with
the values that we calculate from simulations.

If the hot accretion flow is threaded by a strong magnetic field, a
significant fraction of its luminosity could be emitted in the form of
synchrotron radiation which, assuming field strength $B = 
10^4\,G\, \beta^{-1/2}_{10}\; T_{p,12}\;\tau_T^{1/2}\,M_7^{-1/2}$,
could reach 
\begin{eqnarray}
L_{\rm syn} \approx 4\times10^{42}\,{\rm erg\,s^{-1}}\beta^{-1}_{10}\,\tau_{T}^2\; M_7^4 \,,
\label{eq_Lsyn}
\end{eqnarray}
where $\beta = 8\pi p_{\rm th}/B^2 = 10\,\beta_{10}$ is the ratio of
thermal to magnetic pressure in the gas, expected to reach values of
$1-10$ in the central regions of RIAFs \citep{cao11}. The presence of
the softer photons supplied in situ by synchrotron and
bremsstrahlung emission would also give rise to the inverse Compton
radiation of similar magnitude:
\begin{eqnarray}
L_{\rm IC} \approx 2\,L_{\rm soft}\;
T_{p,12}^{-2}\;\tau_{T}\; M_7^3 \,,
%\left(\frac{T_p}{10^{12}\,{\rm K}}\right)^{-2}\tau_{T}\; M_7^3 \,,
\label{eq_Lic}
\end{eqnarray}
where relativistic factors have been evaluated for $v/c\approx 0.3$,
appropriate for the BBH regime close to the merger, and $L_{\rm soft}$
is the luminosity of soft radiation. 

Where the high energy tail of protons reaches the threshold of
$kT_p=100$~MeV an additional high energy process contributes to the
radiative cooling: proton-proton collisions result in copious pion
production, followed by pion decay to two $\gamma$-ray photons, $p + p
\rightarrow p + p + \pi^0 \rightarrow p + p + 2\,\gamma$
\citep{dcw74,ek83,cmt86, mnk97,om03,bbm06}. Following \citet{cmt86},
who calculated the $\gamma$-ray emission from the {\it p-p} collisions of a
thermal distribution of protons in vicinity of a single Kerr black
hole, we estimate
\begin{eqnarray}
L_{\rm {\it pp}} \approx 2-13\times10^{39}\,{\rm erg\,s^{-1}}\;
T_{p,12}\; \tau_{T}^2\; M_7 \,,
%\left(\frac{T_p}{10^{12}\,{\rm K}}\right)\tau_{T}^2\; M_7 \,,
\label{eq_Lpp}
\end{eqnarray}
where the two extreme values correspond to a static and maximally
rotating black hole, respectively. We expect the luminosity in the BBH
system to be closer to the higher value because the gas in the
rotating and dynamic spacetime of the pair of orbiting BHs is very
efficiently shock-heated to 100~MeV. The emission of $\gamma$-rays due
to the pion decay is strongly suppressed in the limit $\tau_{T}\gtrsim
1$ due to the increased cross section of $\gamma$-ray photons to
electron--positron pair production, as well as the increased coupling
between electron and proton plasma, which lowers $T_p$ below the
energy threshold for the pion production \citep{cmt86}. In calculating
luminosities in this section, we assumed the gas to be optically thin
within the Bondi radius, which sets {\it an upper limit} on the gas
density of the hot accretion flow, and thus $L_{\rm brem}$, $L_{\rm
syn}$, $L_{\rm IC}$, and $L_{\rm pp}$.

The spectral energy distribution of these sources would be similar to
a group of low luminosity-AGN to which RIAF models have been applied
\citep[][for example]{nemmen06}. Spectral bands where these emission
mechanisms are expected to peak in the reference frame of the binary
are submillimeter (synchrotron), UV/X-ray (inverse Compton), $\sim
1$~MeV $\gamma$-ray (bremsstrahlung and inverse Compton), and $\sim
20$~MeV (pion decay). Additional components that we do not model in
this work but could also arise and in principle overtake the emission
from the hot gas are the wide-band non-thermal synchrotron emission,
if active and persistent jets are present in the system, as well as
the optical/UV emission associated with the accretion disk that may
encompass the hot flow at larger radii \citep{ho05}.

The bottom panels in Figure~\ref{fig:cloud_brems} focus on
oscillations in luminosity due to relativistic beaming and Doppler
boosting of light emitted by the gas wakes. We account for these
effects by multiplying the broadband emissivity of the gas (i.e., the
luminosity per unit volume) by a factor
$(W(1-\beta\cos(\theta)))^{-4}$ and integrate over the volume to
obtain the bolometric luminosity. Here, $W$ is the local Lorentz
factor and $\beta \cos(\theta)$ is the projection of the local
3-velocity to the line of sight of the observer. We do not account for
the bending of light and gravitational redshift of photons in the
potential well of the BBH. The variations in luminosity shown in
Figure~\ref{fig:cloud_brems} are calculated for a distant observer
placed in the plane of the binary at infinity, an orientation for
which the oscillations in the light curve are maximized. To emphasize
the oscillations, the bottom panels show the ratio of beamed to
unbeamed light curves. The most notable difference among simulated
binary configurations is that the $q=1$ case (CH1) yields sinusoidal
oscillations, while the $q=1/2$ cases (CH2, CH3, and CH4) give rise to
double-peaked oscillations. We also find that in configurations with
arbitrary spin orientation (CH3 and CH4) the sinusoidal peaks exhibit
the largest degree of asymmetry. This is because the binary with
parallel spins (CH2) stirs the surrounding gas more uniformly than
binaries with misaligned spins (CH3 and CH4), which precess about the
original orbital plane.

\begin{figure}[t]
\centering
\includegraphics[width=0.45\textwidth]{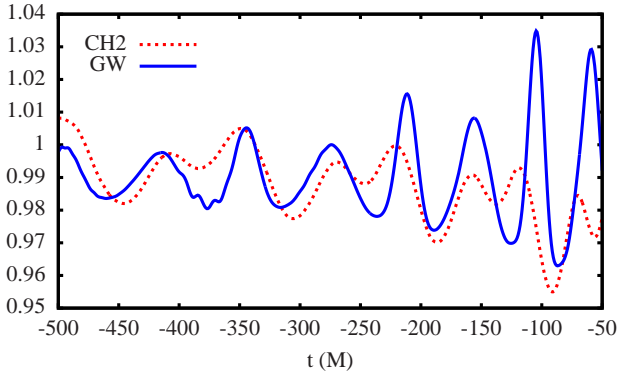}
% the plot* commands are used by the aastex target. Please do not remove them.
%\plottwo{figures/Cloud_GWL-bw}{figures/Cloud_GWL-color}
\caption{Beamed to unbeamed light curve ratio from case CH2 in 
Figure~\ref{fig:cloud_brems} and corresponding GW in arbitrary units.}
\label{fig:gws}
\end{figure}

The bottom-right panel illustrates the dependence of the oscillations
on temperature.  The most prominent oscillations are again associated
with the lower temperature flow, which is more susceptible to shocking
by the binary. While the high temperature of the gas in the CH1 case
prevents formation of the strong density gradients, the lower
temperature flows allow the density wakes to interact and waves of gas
to propagate within the inner region, giving rise to more varied
features in the oscillations. In all cases, the oscillations in
luminosity are mirrored by the GWs.  Figure~\ref{fig:gws} shows how
the double-peaked structure in case CH2 emerges in both the luminosity
and the GWs when the binary is observed edge-on. The double peaks in
the GWs arise from the superposition of $l=2$ and $l=3$ modes in $q
\ne 1$ binaries.

\begin{figure}[t]
\centering
\includegraphics[width=0.45\textwidth]{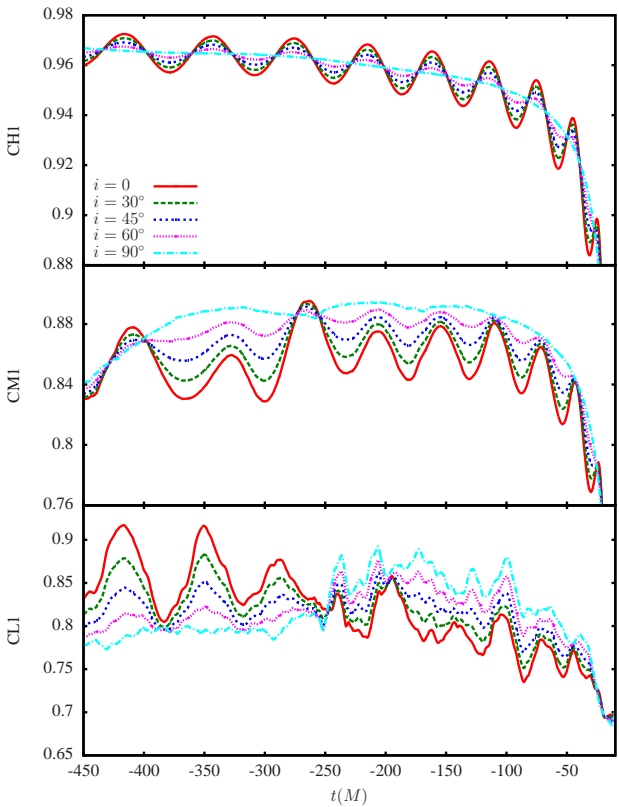}
% the plot* commands are used by the aastex target. Please do not remove them.
%\plottwo{figures/Cloud_VA-bw}{figures/Cloud_VA-color}
\caption{ Variation of the bremsstrahlung luminosity oscillations with
inclination angle $i$ in accretion flows of initial proton temperature
$10^{12}$, $10^{11}$, and $10^{10}$ K from top to bottom,
respectively. }
\label{fig:Lbrem_angled}
\end{figure}

Figure~\ref{fig:Lbrem_angled} shows how oscillations vary as a
function of inclination. The light curves have been evaluated for the
$q=1$ binary for three different temperature cases (CH1, CM1, and CL1)
and the inclination angle is defined with respect to the initial
binary orbital plane. Here again we show the ratio of beamed to
unbeamed light curves. Because the motion of the denser gas wakes is
tied to the plane of the binary, the oscillations in all runs disappear with
increasing inclination angle. The relative drop in the
luminosity of beamed light just prior to the coalescence is a
consequence of de-boosting of light caused by the strong radial inflow
of the gas toward the black holes in this final stage of their
evolution.  The most notable difference among the three runs is that
the oscillation curves exhibit more structure with decreasing
temperature. This can be understood because the lower temperature gas
is more clumpy and more prone to shocks, while in high-temperature
flows, higher thermal velocity of the gas acts to erase density
inhomogeneities. Along similar lines, lower temperature gas has less
pressure support against infall, leading to higher infall velocities
and thus more significant de-boosting with respect to the unbeamed
luminosity case. In the precessing binaries, the time-varying
inclination of the orbital plane imposes an additional modulation of
the oscillations. Of the situations considered in this paper, two
generic BBH systems, CH3 and CH4, precess due to their misaligned
spins. The maximum precession angle attained over the entire evolution
is at most $13^\circ$ with respect to the initial orientation of the
orbital plane, resulting in only minor modulation in their light
curves. 

It is worth noting that accounting for the effects of light
bending and gravitational redshift would result in the modified
variability pattern of emitted light, relative to those shown in
Figure~\ref{fig:Lbrem_angled}. The photons most strongly affected by
the general relativistic effects are emitted by the gas wake
associated with the ``background'' black hole, as seen from the
perspective of an observer. This is because these photons travel
across the deepest part of the potential well of the binary before
escaping to infinity. However, most of the boosted light contributing
to the peaks in the oscillations is emitted by the foreground wake,
associated with the black hole moving toward the observer and thus
with photons which escape from the perimeter of the BBH orbit without
crossing the deepest part of its potential well. These photons will be
less affected by the general relativistic effects, thus partly
justifying our simplistic approach in calculation of the light curves.

\section{Circumbinary Disks}\label{S_disk}

\subsection{Initial Conditions}

\begin{figure*}[ht]
\centering
\includegraphics[width=0.90\textwidth]{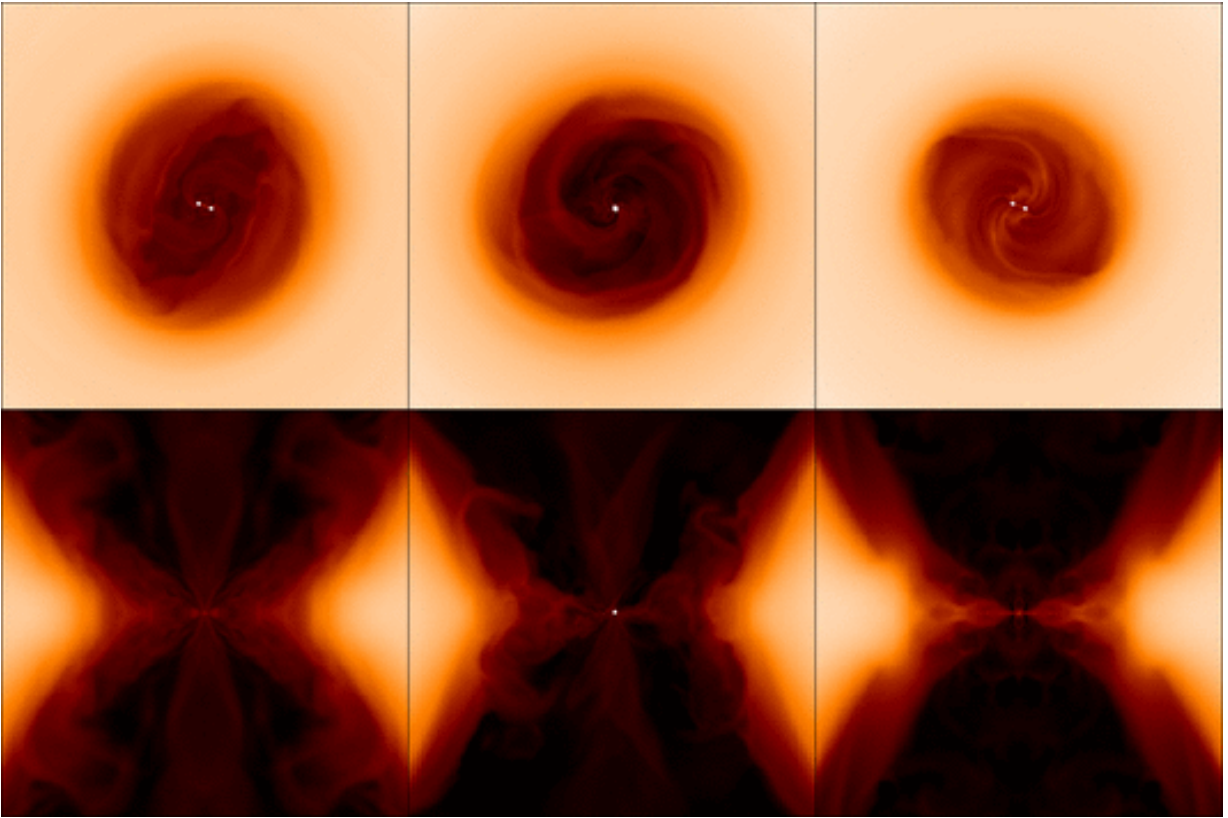}
% the plot* commands are used by the aastex target. Please do not remove them.
%\plotone{figures/Panels2D-bw}\\
%\plotone{figures/Panels2D-color}
\caption{Density distribution of gas in circumbinary disk models DA1,
DA4, and DC1, shown on logarithmic scale, in face-on {\it (top
panels)} and edge-on {\it (bottom panels)} view when the binary
separation is $2M$. White dots mark the positions of the BHs. In panels
where only one or no BHs are visible, this happens because orbital
precession (top middle) or orbital motion (bottom panels) takes BHs
outside of the plane of the image. Panels show a region of size
$60\,M$.}
\label{fig:diskSnaps}
\end{figure*}

In this scenario, we consider circumbinary disks whose inner edge, the
location where the torques from the binary are expected to create a
low-density ``hole'' in the disk center \citep{al94}, is set at $r =
16\,M$.  This choice is motivated by the requirement that the inner
edge of the disk ``follows'' the shrinkage of the binary and remains
at twice its semimajor axis until a binary separation of $8\,M$. In
the context of the \citet{ss73} model of a steady-state accretion
disk, this requirement translates into the value of the half-thickness
ratio of the disk, $h/r$, when the rate of viscous inflow of the disk,
$\dot{a}_{\rm visc} = - 1.5\, (h/r)^2 \,\alpha \,V_K$, is set equal to
the inspiral rate of the binary from gravitational radiation losses
\citep{peters64}. Here, $V_K$ is the Keplerian velocity at the inner
disk edge. This equality is satisfied for $h/r=0.2$ (0.4) and a
viscosity coefficient $\alpha = 0.4$ (0.1). Note that there are
uncertainties associated with this estimate since the steady-state
model does not fully capture the dynamics of the disk close to the BH
coalescence.  We follow \citet{oneill09} and setup gas-pressure-supported
disks characterized by a constant midplane density and
entirely azimuthal initial gas velocities.

As in the case of hot accretion flows, here we only consider BBHs on
quasi-circular orbits. This is based on the expectation that late in
the BBH inspiral any initial orbital eccentricity will be erased due
to the emission of gravitational radiation unless there is a mechanism
which can compete with it. A mechanism that has been suggested to
increase the eccentricity of a BBH orbit is a resonant interaction of
the binary with the circumbinary disk
\citep{an05,cuadra09,roedig11}. It is possible to find the radius at
which the effects of gravitational radiation and circumbinary disk are
competing by setting $\dot{e}$ equal for the two cases. We adopt
$\dot{e}_{\rm visc} = \mathcal{R}\,\dot{a}_{\rm visc}/2a$ and the
standard expression for $\dot{e}_{\rm gw}$ \citep{peters64}. Here, $a$
is the semimajor axis of the binary and $\mathcal{R}= a\,(de/da)$ is
the parameter that relates evolution of the eccentricity to that of
the semimajor axis. \citet{an05} compute $\mathcal{R}$ from
simulations and find $\mathcal{R}\approx-0.1$ for hot disks with
values of $h/r$ comparable to those considered in this work. Assuming
$h/r = 0.2$, $\alpha = 0.4$, $q=1$, and $e = 0.8$, we find that the
two effects balance each other at a binary separation of $a =
118\,M$. The assumed value of the eccentricity follows from a recent
finding that binary eccentricity driven by the disk torques alone
reaches a saturation value in the range 0.6--0.8
\citep{roedig11}. Following the GW evolution of a BBH on an eccentric
orbit with $e = 0.8$ from separation $a = 118\,M$ to $8\,M$, which is
a starting point of our simulations, we find that a maximal residual
eccentricity is $e \approx 0.06$. The residual eccentricity only
weakly depends on the binary mass ratio and has a similar value for
$q=1/2$ binary. This low residual eccentricity is insufficient for the
binary to drive shocks in the distant circumbinary disk and hence, in
the remainder of this paper, we focus our attention on the
quasi-circular BBHs.

The parameters of the circumbinary disk models are summarized in the
bottom half of Table~\ref{tab:big}. All disks are corotating with the
binary except in model DC1 which is retrograde.  To minimize spurious
transients caused by the initial circumbinary disk relaxing to the
dynamic binary potential, the disk is ``relaxed'' for a period of
$\sim 250\,M$. As before, we do not show this initial phase of the
simulations as a part of the results reported here. In the hot
accretion flow scenario this relaxation effectively eliminates
spurious transients. For disks, the relaxation is not as effective and
leaves behind residual slow oscillations in the bulk of the disk.
These do not give rise to transients, and have little effect on the
properties of the gas in the disk hole, where most of the variability
is confined.

\subsection{Properties of the Accretion Flow}

In all simulations a
plunging binary recedes promptly from the inner edge of the disk and
as a consequence, the effect of the binary on the disk beyond its
inner edge is relatively weak. In agreement with \citet{oneill09}, we
do not detect shocks in the body of the disks caused by the binary
motion nor by the mass loss in GWs associated with the coalescence. In
the absence of characteristic variability from the body of the disk, we
focus our discussion throughout the remainder of this section on the
EM counterparts emanating from the disk hole.

The two BHs capture a fraction of the gas from the disk inner edge,
which flows into the central region to form a low-density ambient
medium surrounding the binary. Figure~\ref{fig:diskSnaps} shows the
face-on and edge-on snapshots of the gas density in the central region
of the DA1, DA4, and DC1 cases.  Despite their differences in mass
ratios and spins, the DA1 and DA4 binaries entrain similar amount of
gas from the disk. This is a combination of the two effects: while
$q=1$ on the one hand presents a larger angular momentum barrier for the
inflowing gas, it also inspirals at a slower rate than a $q=1/2$
binary, which allows it to capture gas over a longer period and offset
somewhat the primary effect.  The $q=1$ retrograde disk (DC1), on the
other hand, shows a smaller central disk hole, with a higher gas
density in it relative to the two previous cases. For this case, the
binary and the disk interact to decrease the angular momentum of the
gas, as suggested by \citet{nixon10}, thus allowing a greater inflow
rate.

To quantify the effect of the binary--disk interaction, we calculate
the mean density of the gas, $\rho_{\rm h}$ in the disk hole (i.e., $R
\le 10\,M$) as a fraction of the disk midplane density $\rho_{\rm d}$,
namely, $f_{\rm g} = \rho_{\rm h}/\rho_{\rm d}$.  For $h/r=0.2$ disks,
we find $f_{\rm g} \approx 10^{-5}$, indicating the efficiency of the
binary in suppressing the gas flow into the central region. For the
$h/r=0.4$ disk, $f_{\rm g} \approx 10^{-4}$, owing to the higher
thermal speed of the gas which flows in at a higher rate. The
orientation of spins does not seem to affect the gas inflow rate
significantly.

Similar to the hot flow models the luminosity of the emerging variable
EM signal will be determined by the density of the gas in the vicinity
of the BBH. In our simulations the choice of $h/r$ and $\alpha$
uniquely specify the gas density of the disk in hydrostatic
equilibrium whose vertical structure is supported by the gas thermal
pressure. It is worth noting however that modeling of the gas-pressure-supported
disks in this work is a numerical necessity and that the
innermost regions of realistic accretion disks are expected to be
predominantly supported by radiation pressure \citep{ss73}. A key
difference between the two classes of disks is that for a given
$\alpha$, $M$, and $h/r$ (or equivalently $\dot{M}$) the gas-pressure-supported 
disks are characterized by a higher gas density than
radiation-pressure-supported disks. This can be intuitively understood
because with the radiation as a dominant source of pressure support,
disks need a smaller fraction of the thermal pressure (and hence, gas)
in order to maintain the hydrostatic equilibrium. The implication is
that depending on the dominant emission mechanism, the luminosity of
the gas-pressure-supported disks may not trace that of the realistic
disks for scenarios under consideration. We thus regard the density of
the disk to be a free parameter, which can be scaled to some physical
value, and use $f_{\rm g}$ to estimate the fraction of the gas captured
by the binary. We expect that $f_{\rm g}$ calculated from our
simulations is a robust measure, regardless of the assumed structure
of the disk. This is because the accretion of the gas from the inner
edge of the disk (i.e., its angular momentum transport) is likely
dominated by the binary torques, rather than the radiation pressure or
``viscous'' effects. This point however remains to be confirmed in
future simulations properly equipped to address it.

To gain context about the magnitude of variable EM signal associated
with these systems, let us consider a Shakura--Sunyaev model of a
radiation-pressure-supported, steady-state disk, with $h/r=0.2$ in the
regime where Thomson scattering is the dominant source of opacity
\citep{ss73}. Using maximum radiative efficiency for the Schwarzschild
black hole, one finds at the disk inner edge a temperature $ 2.5\times
10^5$ K and $\rho_{\rm d} \approx 3.5\times 10^{-11}\,{\rm
g\,cm^{-3}}$.  Since in the central region $f_{\rm g} \sim 10^{-5}$
and the gas is promptly shock-heated to $T_p\sim10^{11}$ K, the disk
hole acquires the properties of the hot accretion flow discussed in
the previous section. Assuming that in this case $\varepsilon =
T_e/T_p = 0.1$, the upper limit on the bremsstrahlung luminosity from
the disk hole and Bondi accretion rate onto the BHs are
\begin{eqnarray}
& L_{\rm brem} & \approx  1.6\times 10^{35}\,{\rm erg\,s^{-1}} 
\varepsilon_{-1}^{1/2}\;f_{\rm g,-5}^2\;\rho_{\rm d}^{\prime 2} \; R_{10}^3\; T_{p,11}^{1/2}\;\nonumber\\
 & \times &  \left(1 + 4.4\,\varepsilon_{-1}\,T_{p,11}\right)_{5.4}\; M_7^3 \,,
%  \left(\frac{f_{\rm gas}}{10^{-5}}\right)^2
%  \left(\frac{\rho_{\rm disk}}{3.5\times 10^{-11}{\rm g\, cm^{-3}}}\right)^2 \nonumber\\
%  & \times &  \left(\frac{R}{10\,M}\right)^3 
%  \left(\frac{T_e}{10^{10}\,{\rm K}}\right)^{1/2}
%  \left[ 1 + 4.4\times\left(\frac{T_e}{10^{10}\,{\rm K}}\right) \right]_{5.4} M_7^3 
\label{eq_Lhole}
\end{eqnarray}
\begin{eqnarray}
\dot{M}_{\rm B} & \approx & 2.5\times 10^{-4}\,M_{\odot}\,{\rm yr^{-1}}
f_{\rm g,-5}\; \rho_{\rm d}^\prime\; T_{p,11}^{-3/2}\; M_7^2 \,. 
%\left(\frac{f_{\rm gas}}{10^{-5}}\right)
%\left(\frac{\rho_{\rm disk}}{3.5\times10^{-11} {\rm g\, cm^{-3}}}\right) \nonumber\\
%  & \times &  \left(\frac{T_p}{10^{11}\,{\rm K}}\right)^{-3/2} M_7^2 \; . 
\label{eq_Mhole}
\end{eqnarray}
If a strong magnetic field is present, assuming $\beta=10$,
\begin{eqnarray}
L_{\rm syn} & \approx & 2.5\times10^{36}\,{\rm erg\,s^{-1}}
\beta_{10}^{-1}\; f_{\rm g,-5}^2\;\rho_{\rm d}^{\prime 2} \; R_{10}^3\; T_{p,11}\;M_7^3 \,,
%\beta_{10}^{-1}\,\left(\frac{f_{\rm gas}}{10^{-5}}\right)^2
%\left(\frac{\rho_{\rm disk}}{3.5\times 10^{-11} {\rm g\, cm^{-3}}}\right)^2\nonumber \\
%& &\left(\frac{R}{10M}\right)^3 
% \left(\frac{T_p}{10^{11}\,{\rm K}}\right)  M_7^3 \, ,
\label{eq_Lsynchro} 
\end{eqnarray}
where $\varepsilon = 0.1\varepsilon_{-1}$, $f_{\rm g}= 10^{-5}\,f_{\rm
g,-5}$, $\rho_{\rm d}^\prime = \rho_{\rm d}/3.5\times 10^{-11} {\rm
g}\, {\rm cm^{-3}}$, $R=10M\,R_{10}$, and $T_p = 10^{11}{\rm
K}\,T_{p,11}$. While higher than the bremsstrahlung, the synchrotron
luminosity is still modest in absolute terms for $M\leq 10^7\,\Msun$
BBH systems. As in the case of a hot accretion flow, it can be shown
here that the inverse Compton luminosity can be of order of $L_{\rm
syn}$ and $L_{\rm brem}$. In this case, however, the shocked gas from
the central low-density region can more easily (adiabatically) expand
into the ambient medium. As a consequence, the temperature of the gas
remains below the 100 MeV threshold for $\pi^0$ production in most of
the computational volume and the emission of $\gamma$-ray photons due
to pion decay is expected to be inefficient.

\begin{figure}[t]
\centering
\includegraphics[width=0.45\textwidth]{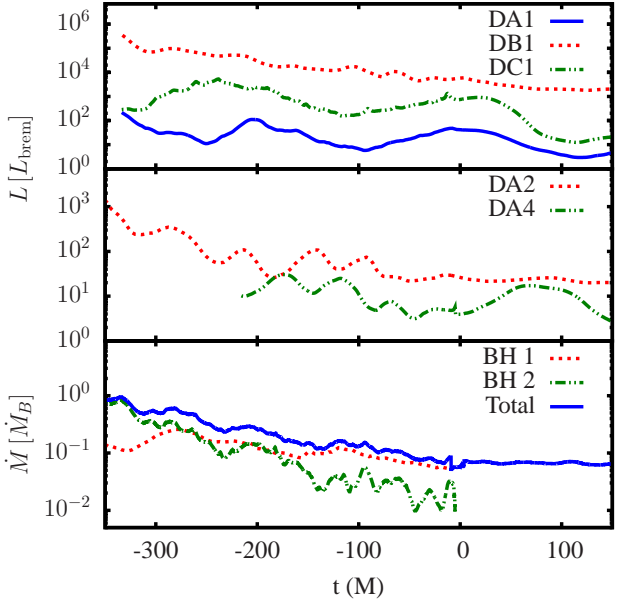}
% the plot* commands are used by the aastex target. Please do not remove them.
%\plottwo{figures/Disks_All-bw}{figures/Disks_All-color}
\caption{Bremsstrahlung luminosity from the disk hole 
(top two panels) in units of $L_{\rm brem}$ given by
Eq. (\ref{eq_Lhole}).  Total and individual BH accretion rates for the
case DA2 (bottom panel) in units of $\dot M_B$ given by
Eq. (\ref{eq_Mhole}).}
\label{fig:diskBrems}
\end{figure}

Figure~\ref{fig:diskBrems} shows the bremsstrahlung light curves
integrated within a sphere of radius $R = 10M$, ensuring contribution
only from the diffuse gas in the disk hole.  The top panel shows the
$q=1$ cases. The long period variations in the luminosity in $h/r =
0.2$ models (DA1 and DC1) correspond precisely to the orbital
frequency at radius $R= 10M$ and arise due to the flow of gas streams
on eccentric orbits in and out of the integration domain.  This is
evident in the first and third panels of Figure~\ref{fig:diskSnaps},
which illustrate that the distribution of gas in the vicinity of $q=1$
binaries is ellipsoidal due to the binary torques. The formation of
the eccentric streams of matter in response to the BBH torques is a
salient property of binary--circumbinary disk systems and has
previously been pointed out in the context of Newtonian simulations
\citep{hayasaki07,mm08,cuadra09}. This type of variation is not
observed in the ``hotter'' $h/r=0.4$ disk (DB1). In this case the
binary is less efficient at driving a resonant motion of the
surrounding gas and it finds itself immersed into a relatively
uniform density medium. The decrease in the overall luminosity in
these three cases is consistent with the scaling $L_{\rm brem} \propto
f_{\rm g}^2$ implied by Eq.~(\ref{eq_Lhole}).  The accretion
rates reflect a similar behavior, namely, the scaling $\dot{M}_{\rm B}
\propto f_{\rm g}$ from Eq.~(\ref{eq_Mhole}).

The middle panel in Figure~\ref{fig:diskBrems} shows light curves from
the $q=1/2$ binaries (DA2 and DA4).  The binaries in these cases
induce a more circularly symmetric distribution of gas within the disk
hole (see DA4 case in Figure~\ref{fig:diskSnaps}).  The variability or
oscillations in the light curves are, unlike for $q=1$ binaries,
associated with the orbiting wakes following the BHs and thus follow
the BBH orbital motion. Notice also that here again the luminosity of
the misaligned spins case (DA4) falls below its counterpart with
parallel spins (DA2). However, unlike the case of the hot accretion
flows there is no pre-merger flare and subsequent drop-off in
luminosity at merger. The lack of these two signatures in the disk
scenario can be attributed to the contrasting density distributions in
the two scenarios: in the hot flow the gas density increases toward
the center of the binary potential, while in the circumbinary disk
case the density is lowest at the center of the system, as illustrated
in the bottom panels of Figure~\ref{fig:diskSnaps}. This occurs
because the gas is only slowly ``leaking'' from the inner edge of the
disk into the central region. The luminosity of a dilute gas powered
by shocks from the BBH motion just before the merger is overtaken by
the luminosity of the remainder of the gas in the disk
hole. Consequently, the luminosity peak associated with the BBH merger
is invisible in the light curves shown in the top panels of
Figure~\ref{fig:diskBrems}.

The bottom panel in Figure~\ref{fig:diskBrems} shows the accretion
rates onto the primary and secondary BHs, as well as the total, in
model DA2. Note that initially the accretion rate onto the lower mass
secondary BH exceeds that of the primary, because it orbits closer to
the inner edge of the disk where it captures and accretes gas
\citep{al96,gould00}. Later in the inspiral, the primary increasingly
captures gas bound to the secondary and augments its accretion rate
until time $\approx -250M$ before merger, when the accretion rates of
the primary and secondary are reversed and the primary becomes the
dominant accretor.

Our simulations indicate that low-luminosity variability described in
this section would be challenging to detect in observational campaigns
searching for BBHs, except perhaps for massive systems with $M\gg
10^7\,\Msun$. The situation is more severe for ``thinner'' disks in
which the central regions are expected to be even less luminous. In
these disks, the radial inflow velocity at the inner edge is $\propto
(h/r)^2$, resulting in the decoupling of the binary from the disk even
earlier in the inspiral. As a consequence, the gas in the disk hole
will have lower density and thus be even dimmer.  The circumbinary
disk itself will likely overwhelm the overall emission and shield any
sign of variability from the disk hole although, strictly speaking,
the light emitted by the disk is expected to peak in the optical and
UV-band, while the bremsstrahlung and inverse Compton emission from
the $T_p\sim10^{11}{\rm K}$ gas in the hole is expected to peak at
$\sim 100$\,keV. Because of its radiatively inefficient nature the gas
in the disk hole would exhibit similar emission properties to the hot
flow described in \S~\ref{S_properties1}.

%%%%%%%%%%%%%%%%%%%%%%%%%%%%%%%%%%%%%%%%%%%%%%%%%%%%%%%%%%%%%%%%%%%%%%%%

\section{Conclusions}\label{S_conclusions}

We studied the EM signatures associated with the late inspiral and
merger of supermassive BBHs with unequal masses, different BH spin
orientations, and two environments (hot accretion flows and
circumbinary disks).

In a \emph{hot accretion flow}, the plunge and merger of the binary give
rise to a characteristic flare followed by a sudden drop-off. This and
earlier hydrodynamic simulations \citep{bode10,farris10} indicate that
the flare is a robust signature that arises in all modeled BBH
configurations and hot accretion flows regardless of the specific
setup of initial conditions. The amplitude of the flare and the
drop-off are more pronounced in lower temperature flows, where
shocking is more predominant. We find that the shape and amplitude of
the flare depend on a mass ratio and spin configuration to a lesser
degree. A hot flow is also characterized by quasi-periodic variability
from the beaming of light emitted by the gas wakes behind the BHs. In
all cases, this EM equivalent of a ``chirp'' is directly mirrored by
the GWs \citep[existence of such signature has previously been
predicted by][]{kocsis08}.  If observed, either signature (flare or
variability) would be a smoking gun for an impending BBH coalescence
in a hot accretion flow environment.

In the {\it circumbinary disk} case we modeled geometrically thick disks
characterized by high radial inflow velocities, which can follow the
binary until late in the GW-driven inspiral. We find that a
combination of the properties of such disks and the prompt decoupling of
the binary from their inner edge result in an absence of shocks and
luminous EM flares from the body of the disks. The only region where
variable EM signatures emerge is the central low-density hole
surrounding the binary. We find that the gas in this central region
also has the properties of a radiatively inefficient accretion
flow. In unequal-mass BBH systems, its luminosity exhibits variability
related to the BBH orbital dynamics. In equal-mass systems however,
the variability traces the dynamics of the gas streams plunging from
the inner edge of the disk and is not correlated with the frequency of
GWs. In all cases, the emission from the central region is low and
based on our models, unlikely to be observed for BBH systems $\leq
10^7\Msun$.

In summary, while BBH mergers in both the hot accretion flows and
circumbinary disks exhibit characteristic EM signatures, we find that
the former is more likely to be sufficiently luminous to be observed.
To be observable, any variability must be more pronounced than the
natural variability of "normal" AGNs. While the flare signaling the
merger seems to satisfy this condition, the quasi-periodic
oscillations will be more challenging to observe. A precursor GW
detection by a space interferometer could in principle alert EM
observatories in advance, and hopefully alleviate this challenge.

An interesting question is whether, in the absence of a GW precursor,
the flares and variability can be detected in stand-alone EM
observations. Given that more massive BBH systems are also expected to
be more luminous, a future serendipitous discovery of $\gtrsim
10^8\Msun$ BBH coalescence by a burst alert mission cannot be
excluded. A more systematic search will require deep monitoring of the
transient sky with multiwavelength synoptic sky surveys. While this
biases EM searches toward the high BH masses, they are complimentary
to future GW observations that will likely be optimized to search for
the lower mass end supermassive BBHs. Both will be required in order
to eventually understand the properties of the BBH population and
their role in evolution of galaxies and structure in the universe.

%%%%%%%%%%%%%%%%%%%%%%%%%%%%%%%%%%%%%%%%

\acknowledgements We thank the anonymous referee for thoughtful
comments which helped to improve this manuscript. T. Bogdanovi\'c is
supported by a NASA Einstein Postdoctoral Fellowship Award PF9-00061
from Chandra X-ray Observatory Center operated by Smithsonian
Astrophysical Observatory for and on behalf of NASA contract
NAS8-03060. This work is supported by NSF grants 0653443, 0855892, 0914553,
0941417, 0903973, and 0955825. Computations at Teragrid TG-MCA08X009 and
Georgia Tech FoRCE cluster.

%%%%%%%%%%%%%%%%%%%%%%%%%%%%%%%%%%%%%%%%

\bibliographystyle{apj}

\begin{thebibliography}{45}
\expandafter\ifx\csname natexlab\endcsname\relax\def\natexlab#1{#1}\fi

\bibitem[{Ansorg {et~al.}(2004)Ansorg, Br{\"u}gmann, \& Tichy}]{Ansorg:2004ds}
Ansorg, M., Br{\"u}gmann, B., \& Tichy, W. 2004, Phys. Rev. D, 70, 064011

\bibitem[{{Armitage} \& {Natarajan}(2002)}]{an02}
{Armitage}, P.~J., \& {Natarajan}, P. 2002, Astrophys. J., 567, L9

\bibitem[{{Armitage} \& {Natarajan}(2005)}]{an05}
---. 2005, \apj, 634, 921

\bibitem[{{Artymowicz} \& {Lubow}(1994)}]{al94}
{Artymowicz}, P., \& {Lubow}, S.~H. 1994, Astrophys. J., 421, 651

\bibitem[{{Artymowicz} \& {Lubow}(1996)}]{al96}
---. 1996, \apjl, 467, L77+

\bibitem[{Baiotti {et~al.}(2005)Baiotti, Hawke, Montero, L{\"o}ffler, Rezzolla,
  Stergioulas, Font, \& Seidel}]{Baiotti:2004wn}
Baiotti, L., Hawke, I., Montero, P.~J., L{\"o}ffler, F., Rezzolla, L.,
  Stergioulas, N., Font, J.~A., \& Seidel, E. 2005, Phys. Rev. D, 71, 024035

\bibitem[{{Bhattacharyya} {et~al.}(2006){Bhattacharyya}, {Bhatt}, \&
  {Misra}}]{bbm06}
{Bhattacharyya}, S., {Bhatt}, N., \& {Misra}, R. 2006, \mnras, 371, 245

\bibitem[{{Bode} {et~al.}(2010){Bode}, {Haas}, {Bogdanovi{\'c}}, {Laguna}, \&
  {Shoemaker}}]{bode10}
{Bode}, T., {Haas}, R., {Bogdanovi{\'c}}, T., {Laguna}, P., \& {Shoemaker}, D.
  2010, Astrophys.~J., 715, 1117

\bibitem[{{Bogdanovi{\'c}} {et~al.}(2008){Bogdanovi{\'c}}, {Smith},
  {Sigurdsson}, \& {Eracleous}}]{bogdanovic08}
{Bogdanovi{\'c}}, T., {Smith}, B.~D., {Sigurdsson}, S., \& {Eracleous}, M.
  2008, \apjs, 174, 455

\bibitem[{Cactus(2010)}]{Cactuscode.org:web}
Cactus. 2010, {Cactus} Computational Toolkit

\bibitem[{{Cao}(2011)}]{cao11}
{Cao}, X. 2011, \apj, 737, 94

\bibitem[{{Colpi} {et~al.}(2007){Colpi}, {Dotti}, {Mayer}, \&
  {Kazantzidis}}]{colpi07}
{Colpi}, M., {Dotti}, M., {Mayer}, L., \& {Kazantzidis}, S. 2007,
  arXiv:0710.5207

\bibitem[{{Colpi} {et~al.}(1986){Colpi}, {Maraschi}, \& {Treves}}]{cmt86}
{Colpi}, M., {Maraschi}, L., \& {Treves}, A. 1986, \apj, 311, 150

\bibitem[{{Cuadra} {et~al.}(2009){Cuadra}, {Armitage}, {Alexander}, \&
  {Begelman}}]{cuadra09}
{Cuadra}, J., {Armitage}, P.~J., {Alexander}, R.~D., \& {Begelman}, M.~C. 2009,
  \mnras, 393, 1423

\bibitem[{{Dahlbacka} {et~al.}(1974){Dahlbacka}, {Chapline}, \&
  {Weaver}}]{dcw74}
{Dahlbacka}, G.~H., {Chapline}, G.~F., \& {Weaver}, T.~A. 1974, \nat, 250, 36

\bibitem[{{Dotti} {et~al.}(2006){Dotti}, {Salvaterra}, {Sesana}, {Colpi}, \&
  {Haardt}}]{dotti06}
{Dotti}, M., {Salvaterra}, R., {Sesana}, A., {Colpi}, M., \& {Haardt}, F. 2006,
  \mnras, 372, 869

\bibitem[{{Eilek} \& {Kafatos}(1983)}]{ek83}
{Eilek}, J.~A., \& {Kafatos}, M. 1983, \apj, 271, 804

\bibitem[{{Einstein Toolkit}(2010)}]{EinsteinToolkit.org:web}
{Einstein Toolkit}. 2010, {Einstein Toolkit}: Open software for relativistic
  astrophysics

\bibitem[{{Farris} {et~al.}(2010){Farris}, {Liu}, \& {Shapiro}}]{farris10}
{Farris}, B.~D., {Liu}, Y.~T., \& {Shapiro}, S.~L. 2010, Phys.~Rev.~D, 81,
  084008

\bibitem[{{Gould} \& {Rix}(2000)}]{gould00}
{Gould}, A., \& {Rix}, H. 2000, \apjl, 532, L29

\bibitem[{{Hayasaki} {et~al.}(2007){Hayasaki}, {Mineshige}, \&
  {Sudou}}]{hayasaki07}
{Hayasaki}, K., {Mineshige}, S., \& {Sudou}, H. 2007, \pasj, 59, 427

\bibitem[{{Ho}(2005)}]{ho05}
{Ho}, L.~C. 2005, \apss, 300, 219

\bibitem[{Husa {et~al.}(2006)Husa, Hinder, \& Lechner}]{Husa:2004ip}
Husa, S., Hinder, I., \& Lechner, C. 2006, Comput. Phys. Commun., 174, 983

\bibitem[{{Ichimaru}(1977)}]{ichimaru77}
{Ichimaru}, S. 1977, Astrophys.~J., 214, 840

\bibitem[{{Klein} {et~al.}(2009){Klein}, {Jetzer}, \&
  {Sereno}}]{2009PhRvD..80f4027K}
{Klein}, A., {Jetzer}, P., \& {Sereno}, M. 2009, \prd, 80, 064027

\bibitem[{{Kocsis} {et~al.}(2008){Kocsis}, {Haiman}, \& {Menou}}]{kocsis08}
{Kocsis}, B., {Haiman}, Z., \& {Menou}, K. 2008, \apj, 684, 870

\bibitem[{{MacFadyen} \& {Milosavljevi{\'c}}(2008)}]{mm08}
{MacFadyen}, A.~I., \& {Milosavljevi{\'c}}, M. 2008, \apj, 672, 83

\bibitem[{{Mahadevan} {et~al.}(1997){Mahadevan}, {Narayan}, \&
  {Krolik}}]{mnk97}
{Mahadevan}, R., {Narayan}, R., \& {Krolik}, J. 1997, \apj, 486, 268

\bibitem[{{Milosavljevi{\'c}} \& {Phinney}(2005)}]{mp05}
{Milosavljevi{\'c}}, M., \& {Phinney}, E.~S. 2005, Astrophys.~J., 622, L93

\bibitem[{{M{\"o}sta} {et~al.}(2010){M{\"o}sta}, {Palenzuela}, {Rezzolla},
  {Lehner}, {Yoshida}, \& {Pollney}}]{mosta10}
{M{\"o}sta}, P., {Palenzuela}, C., {Rezzolla}, L., {Lehner}, L., {Yoshida}, S.,
  \& {Pollney}, D. 2010, \prd, 81, 064017

\bibitem[{{Narayan} \& {Yi}(1994)}]{ny94}
{Narayan}, R., \& {Yi}, I. 1994, Astrophys.~J., 428, L13

\bibitem[{{Nemmen} {et~al.}(2006){Nemmen}, {Storchi-Bergmann}, {Yuan},
  {Eracleous}, {Terashima}, \& {Wilson}}]{nemmen06}
{Nemmen}, R.~S., {Storchi-Bergmann}, T., {Yuan}, F., {Eracleous}, M.,
  {Terashima}, Y., \& {Wilson}, A.~S. 2006, \apj, 643, 652

\bibitem[{{Nixon} {et~al.}(2011){Nixon}, {Cossins}, {King}, \&
  {Pringle}}]{nixon10}
{Nixon}, C.~J., {Cossins}, P.~J., {King}, A.~R., \& {Pringle}, J.~E. 2011,
  \mnras, 412, 1591

\bibitem[{{Oka} \& {Manmoto}(2003)}]{om03}
{Oka}, K., \& {Manmoto}, T. 2003, \mnras, 340, 543

\bibitem[{{O'Neill} {et~al.}(2009){O'Neill}, {Miller}, {Bogdanovi{\'c}},
  {Reynolds}, \& {Schnittman}}]{oneill09}
{O'Neill}, S.~M., {Miller}, M.~C., {Bogdanovi{\'c}}, T., {Reynolds}, C.~S., \&
  {Schnittman}, J.~D. 2009, \apj, 700, 859

\bibitem[{{Palenzuela} {et~al.}(2009){Palenzuela}, {Anderson}, {Lehner},
  {Liebling}, \& {Neilsen}}]{palenzuela09}
{Palenzuela}, C., {Anderson}, M., {Lehner}, L., {Liebling}, S.~L., \&
  {Neilsen}, D. 2009, Phys.~Rev.~Lett., 103, 081101

\bibitem[{{Palenzuela} {et~al.}(2010){Palenzuela}, {Lehner}, \&
  {Liebling}}]{palenzuela10}
{Palenzuela}, C., {Lehner}, L., \& {Liebling}, S.~L. 2010, Science, 329, 927

\bibitem[{{Peters}(1964)}]{peters64}
{Peters}, P.~C. 1964, Phys.~Rev., 136, 4B

\bibitem[{{Roedig} {et~al.}(2011){Roedig}, {Dotti}, {Sesana}, {Cuadra}, \&
  {Colpi}}]{roedig11}
{Roedig}, C., {Dotti}, M., {Sesana}, A., {Cuadra}, J., \& {Colpi}, M. 2011,
  \mnras, 415, 3033

\bibitem[{Schnetter {et~al.}(2004)Schnetter, Hawley, \&
  Hawke}]{Schnetter:2003rb}
Schnetter, E., Hawley, S.~H., \& Hawke, I. 2004, Class. Quantum Grav., 21, 1465

\bibitem[{{Shakura} \& {Sunyaev}(1973)}]{ss73}
{Shakura}, N.~I., \& {Sunyaev}, R.~A. 1973, Astron. \& Astrophys., 24, 337

\bibitem[{{Sharma} {et~al.}(2007){Sharma}, {Quataert}, {Hammett}, \&
  {Stone}}]{sharma07}
{Sharma}, P., {Quataert}, E., {Hammett}, G.~W., \& {Stone}, J.~M. 2007, \apj,
  667, 714

\bibitem[{Thornburg(2004)}]{Thornburg:2003sf}
Thornburg, J. 2004, Class. Quantum Grav., 21, 743

\bibitem[{Trias \& Sintes(2008)}]{PhysRevD.77.024030}
Trias, M., \& Sintes, A.~M. 2008, Phys. Rev. D, 77, 024030

\bibitem[{{van Meter} {et~al.}(2010){van Meter}, {Wise}, {Miller}, {Reynolds},
  {Centrella}, {Baker}, {Boggs}, {Kelly}, \& {McWilliams}}]{vanmeter09}
{van Meter}, J.~R., {et~al.} 2010, Astrophys.~J., 711, L89

\end{thebibliography}

\end{document}